\documentclass[final,3p,times,twocolumn]{elsarticle}

\usepackage{amssymb}

\newcommand{\cuba}{Cu$_{13}$Ba}

\journal{Journal of Alloys and Compounds}

\begin{document}

\begin{frontmatter}

\title{Single crystal growth and characterization of the large-unit-cell compound Cu$_{13}$Ba}

\author[1]{A. Jesche}
\ead{jesche@ameslab.gov}
\author[1,2]{S. L. Bud'ko}
\author[1,2]{P. C. Canfield}
\address[1]{The Ames Laboratory, Iowa State University, Ames, USA}
\address[2]{Department of Physics and Astronomy, Iowa State University, Ames, USA}

\begin{abstract}
Single crystals of \cuba~were successfully grown out of Ba-Cu self flux. 
Temperature dependent magnetization, $M(T)$, electrical resistivity, $\rho(T)$, and specific heat, $C_{\rm p}(T)$, data are reported.
Isothermal magnetization measurements, $M(H)$, show clear de Haas-van Alphen oscillations at $T = 2$\,K for applied fields as low as $\mu_0H = 1$\,T. 
An anomalous behavior of the magnetic susceptibility is observed up to $T \approx 50$\,K reflecting the effect of de Haas-van Alphen oscillations at fairly high temperatures.
The field- and temperature-dependencies of the magnetization indicate the presence of diluted magnetic impurities with a concentration of the order of 0.01\,at.\%. 
Accordingly, the minimum and lower temperature rise observed in the electrical resistivity at and below $T = 15$\,K is attributed to the Kondo-impurity effect.
\end{abstract}


\end{frontmatter}

\section{Introduction}
Ordered, stoichiometric compounds consisting primarily of one element, e.g., over 90\% elemental, necessarily have many atoms per formula unit and crystallize with large unit cells. 
They are often very sensitive to small changes in composition and even dilute additions can dominate the structural and physical properties\,\cite{Jia2007}. 
Furthermore, they can be interesting candidates to search for quasi-crystals close to crystalline approximants\,\cite{Canfield2010,Goldman2013}.
\cuba~consists of 93\,at.\% elemental copper and crystallizes in a cubic unit cell with a lattice parameter of 11.7\,\AA. 
The strong influence of the 7\,at.\% barium is already apparent from the large reduction of the melting temperature of pure copper ($T_{\rm m} = 1085^\circ$C) to the peritectic decomposition temperature of \cuba~($T_{\rm decomp} = 758^\circ$C)\,\cite{Okamoto1998}. 
The deep eutectic in the binary alloy phase diagram (Fig.\,\ref{phasediagram} after Ref.\,\cite{Okamoto1998}) allows for crystal growth at $T < 700^\circ$C which can be easily achieved by standard laboratory equipment.

In this paper we present structural, thermodynamic and transport data on solution grown single crystals of \cuba. 
We find that it forms as a metal with a moderately low $\gamma < 2$\,mJ/(K$^{2}$\,mol-atomic) and with a quality that allows clear measurement of de Haas-van Alphen oscillations at temperatures of 10\,K in applied fields as low as 2\,T.

\section{Experimental}
Laue-back-reflection patterns were taken with an MWL-110 camera manufactured by Multiwire Laboratories. 
X-ray powder diffraction (XRPD) was performed on ground single crystals using a Rigaku Miniflex diffractometer (Cu-$K\alpha_{1,2}$ radiation).
Magnetization measurements were performed using a Quantum Design Magnetic Property Measurement System equipped with a 7 Tesla magnet.
Electrical resistivity was measured in 4-point geometry using the AC transport option of a Quantum Design Physical Property Measurement System (PPMS). 
The samples were cleaved into rectangular prisms with an aspect ratio of 2-3. Sample geometry and size of the electrical contacts cause an error of $\sim30$\,\% for the absolute value of the electrical resistivity.
EPO-TEK H20E Silver epoxy was used to make electrical contacts on the samples which were cured at $T = 120^\circ$C under air for $\approx$\,20\,min. 
The samples did not visually degrade and XRPD measurements revealed no changes in the diffraction pattern as a result of this treatment.

\section{Crystal growth}

\begin{figure}
\center
\includegraphics[width=0.44\textwidth]{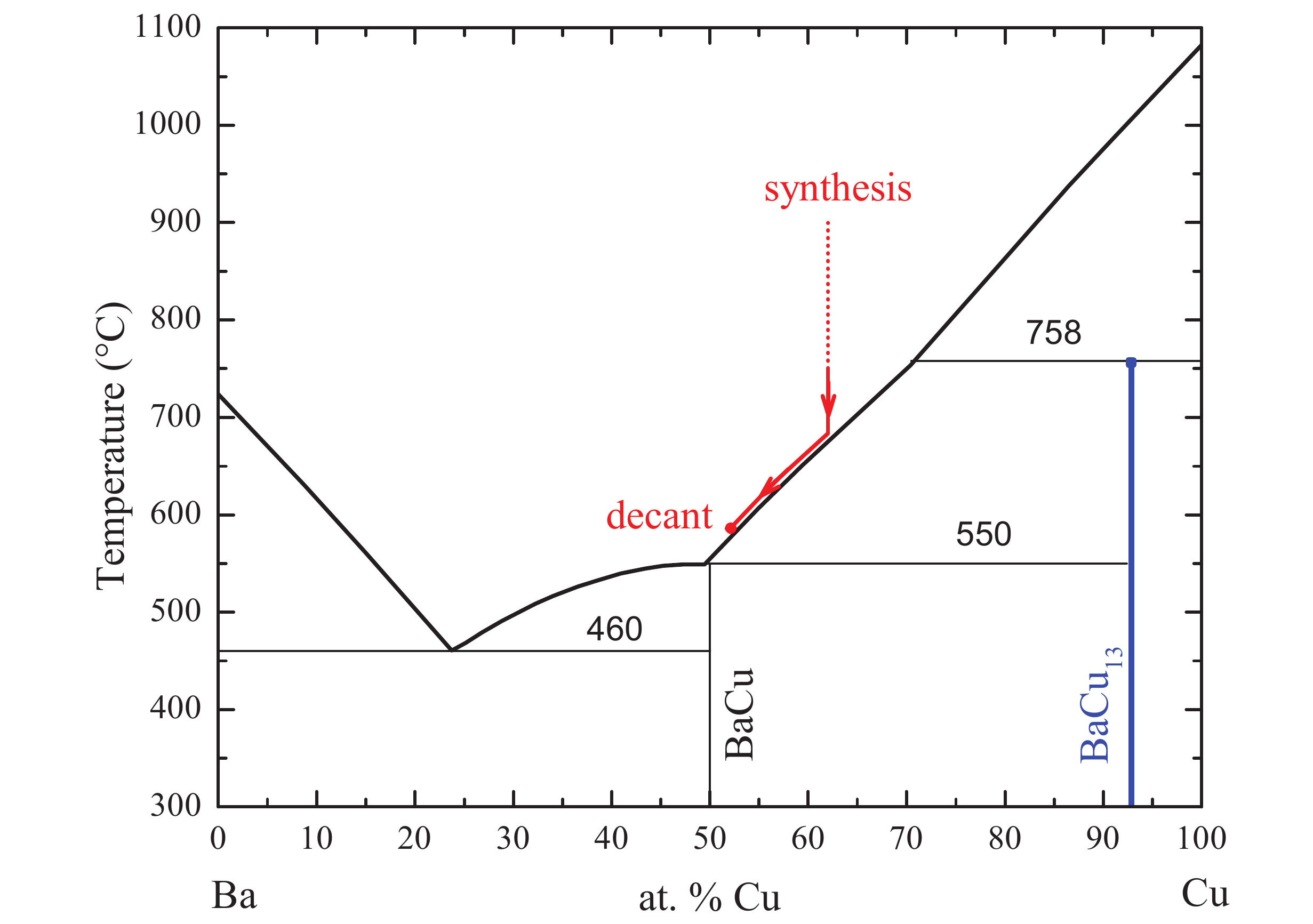}
\caption{Ba-Cu phase diagram after Okamoto \textit{et al.}\,\cite{Okamoto1998}. Starting composition and temperature profile for \cuba~growths are shown with the red line (the dotted line represents rapid cooling). 
}
\label{phasediagram}
\end{figure}
Ba and Cu were mixed in a molar ratio of 38:62 motivated by the published phase diagram\,\cite{Okamoto1998} (Fig.\,\ref{phasediagram}) with a total mass of roughly 2.5\,g.
Ta and alumina were tested as crucible materials and were both found suitable with no significant differences in structural or physical properties of the \cuba-samples.
The 3-cap Ta-crucible\,\cite{Canfield2001} was packed in an Ar-filled glove box and sealed by arc-melting under 0.5\,bar Ar.
The alumina crucible was packed in an Ar-filled glove box with a second catch crucible stuffed with silica wool on top of the growth crucible\cite{Canfield1992}.
The packed crucibles were sealed in a silica ampule under 1/3\,bar Ar. 

The Ba-Cu mixture was heated to $T = 900^\circ$C over 4\,h, cooled to $T = 750^\circ$C within 1.5\,h, slowly cooled to $T = 580^\circ$C over 51\,h and finally decanted to separate the \cuba\,crystals from the excess liquid. A second attempt was performed using a similar temperature profile with holding the temperature at $900^\circ$C for 3\,h and slower cooling from $750^\circ$C to $580^\circ$C over 108\,h (in an alumina crucible). 

\begin{figure}
\center
\includegraphics[width=0.35\textwidth]{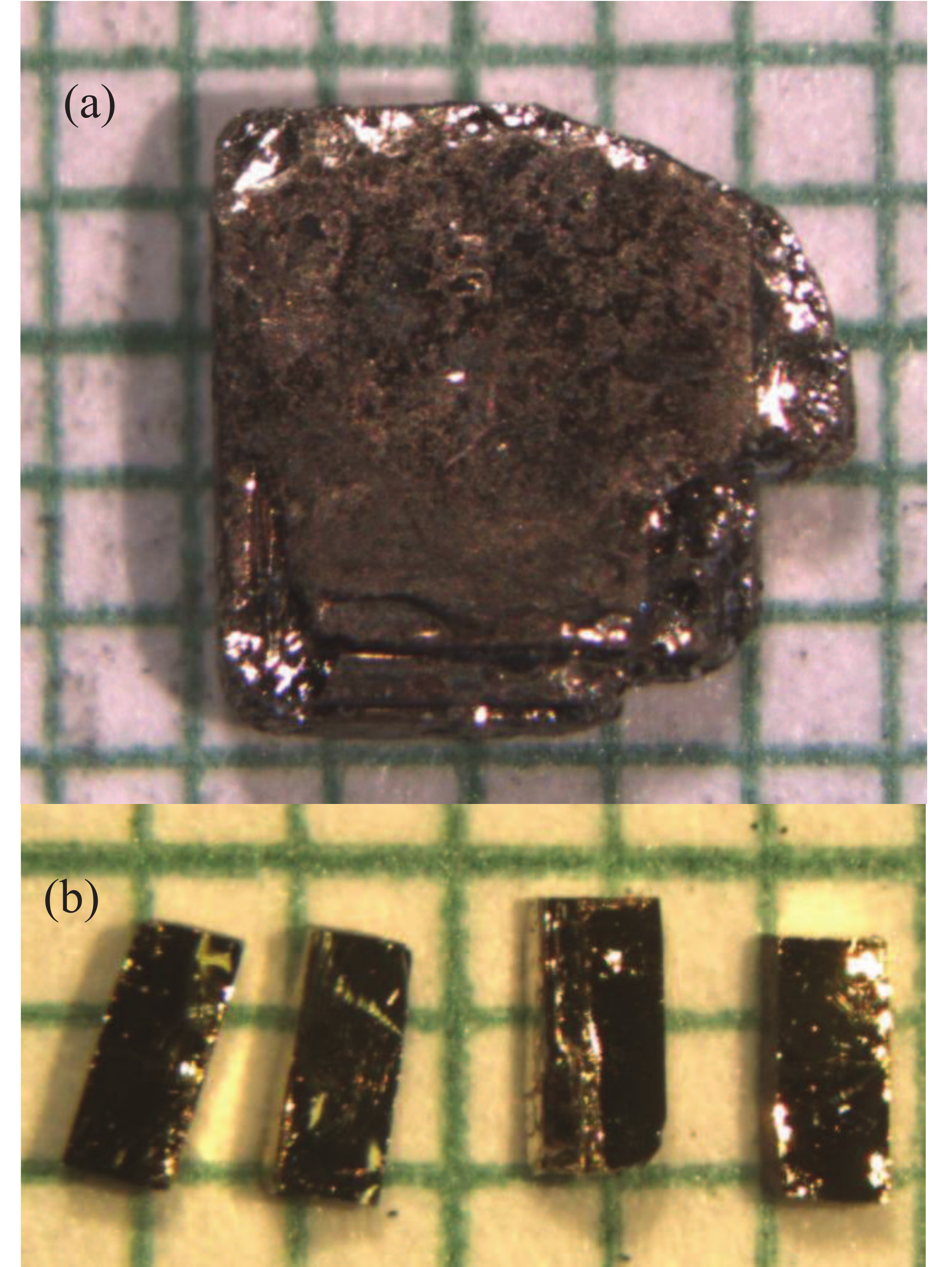}
\caption{(a) As-grown \cuba~single crystal on a millimeter grid. The crystal is 2\,mm tall. (b) Single crystals cleaved with a razor blade show a metallic luster and keep it for several hours in air.
}
\label{sc}
\end{figure}

Single crystals of cubic habit with dimensions up to 5\,mm could be obtained (Fig.\,\ref{sc}a). The crystals cleave well along high symmetry directions and fresh surfaces keep a metallic luster for several hours when exposed to air (Fig.\,\ref{sc}b).

\section{Structural aspects}

\begin{figure}[t]
\center
\includegraphics[width=0.4\textwidth]{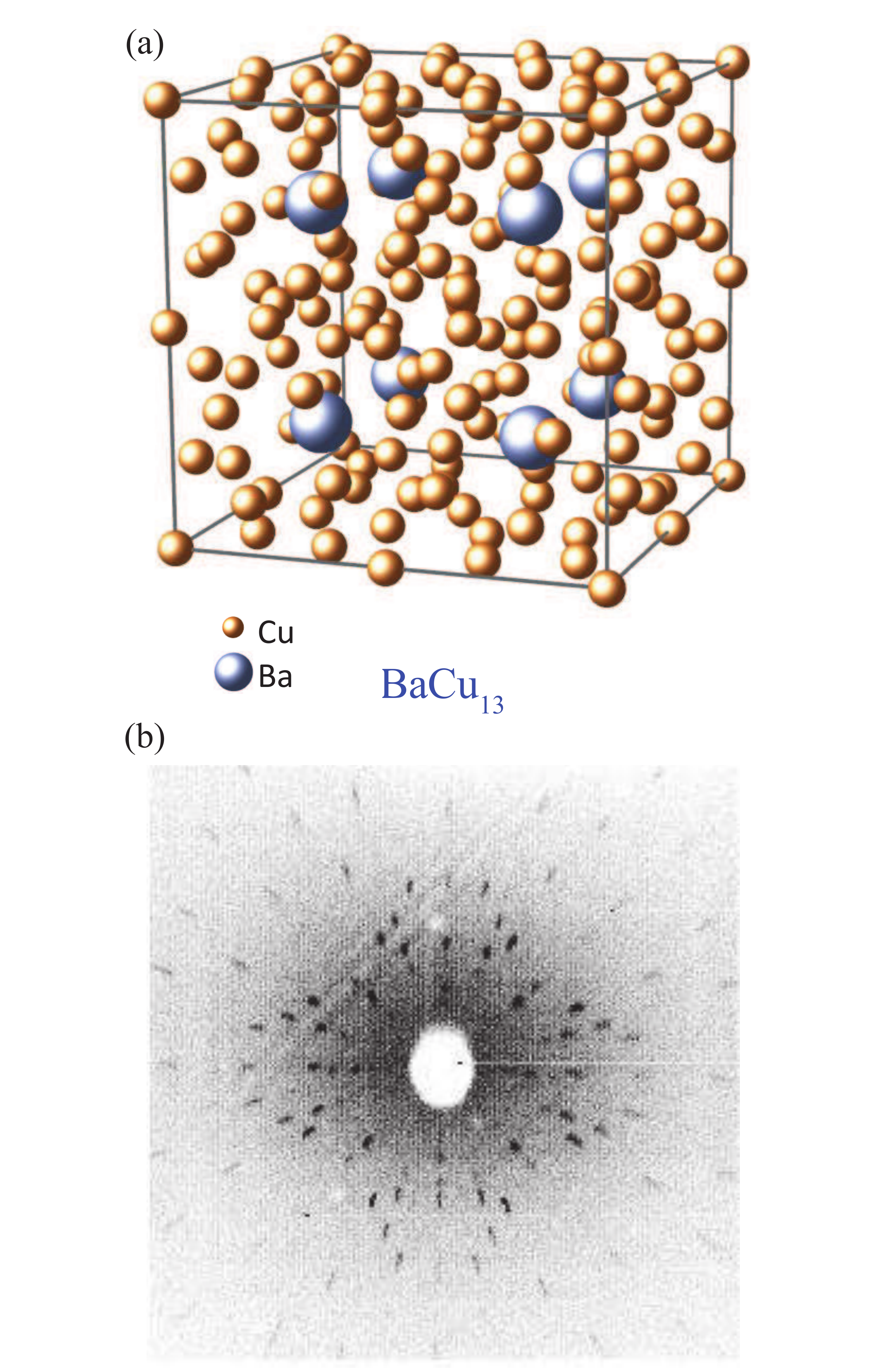}
\caption{(a) Face-centred cubic unit cell of \cuba.
(b) Laue-back-reflection pattern confirming that the single crystals show predominantly $\{1\,0\,0\}$ faces and cleave along the corresponding planes.  
}
\label{unit-cell_laue}
\end{figure}

\cuba~crystallizes in the cubic, NaZn$_{13}$, structure type (Fig.\,\ref{unit-cell_laue}a, space group $F\,m\,\bar{3}\,c$)\,\cite{Braun1959}.
The facets show a four fold rotation symmetry corresponding to \{1\,0\,0\} planes of the cubic lattice as confirmed by Laue back reflection (Fig.\,\ref{unit-cell_laue}b).

\begin{figure}
\center
\includegraphics[width=0.44\textwidth]{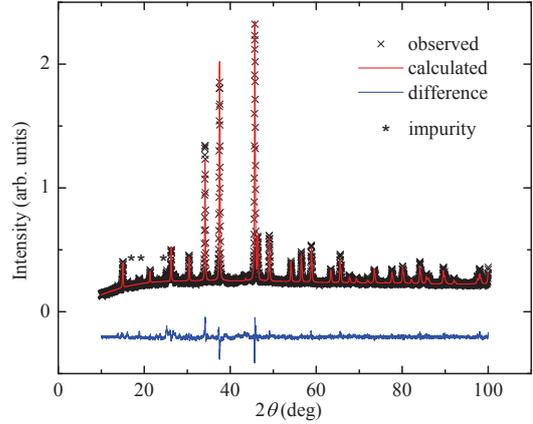}
\caption{Powder X-ray-diffraction pattern of \cuba~with fits based on the published crystal structure. A small amount of an impurity phase, presumably a barium hydroxide hydrate, is inferred from additional peaks marked by asterisks.}
\label{diff}
\end{figure}

The X-ray powder diffraction measured on ground single crystals is shown in Fig.\,\ref{diff}. The diffraction pattern did not change after keeping the powder exposed to air for several days. A Rietveld refinement was carried out using GSAS\,\cite{Larson2000} and EXPGUI\,\cite{Toby2001}.
Instrument parameters for profile function 3 were determined prior to the measurement using a Si-standard: 
only lattice parameter, sample displacement, transparency, and Lorentzian coefficients were released - all other parameters (except background and scaling) were kept constant during the refinement.
In this way, a weighted profile $R$-factor of $R_{\rm wp} = 4.7$ was achieved.
The lattice parameter $a = 11.73(1)$\,\AA~is in good agreement with the literature data ($a = 11.719$\,\AA\,\cite{Braun1959}, $a = 11.754(2)$\,\AA\,\cite{Bruzzone1971}, and $a = 11.697(8)$\,\AA\,\cite{Wendorff2006}).

A small amount of an impurity phase is inferred from the diffraction pattern but the low intensity does not allow for an unambiguous identification. Based on the peak positions it is likely a Ba hydroxide hydrate. This phase can be a decay product of \cuba~or, more likely, of Ba-rich undecanted flux remnants from the sample surface.  

\section{Electrical resistivity}
\begin{figure}[t]
\center
\includegraphics[width=0.44\textwidth]{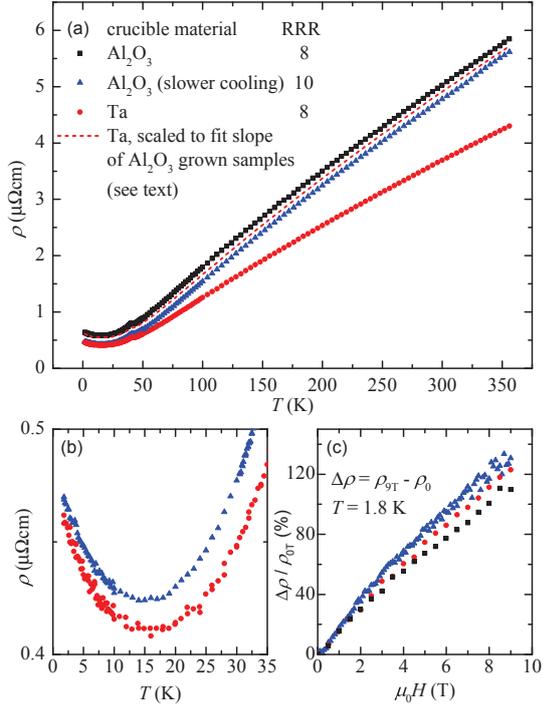}
\caption{Electrical resistivity of \cuba~for three different samples. Crucible material and cooling rate of the crystal growth have only minor effects. 
(a) The linear temperature dependence of a metal is observed over a wide temperature range. 
A small peak at $T \approx 40$\,K is an artifact due to distorted signal readback in the PPMS\,\cite{qd2000}. 
(b) The minimum at $T = 15$\,K is typical for a single-ion Kondo-effect and is in agreement with the small local-moment-contribution inferred from the magnetization measurements.
(c) The magnetoresistance shows an unusual, sublinear field-dependence with a change of slope at $\mu_0H = 3$\,T (inset).
}
\label{res}
\end{figure}

Figure\,\ref{res} shows the temperature-dependent electrical resistivity of \cuba~with current flow along $[1\,0\,0]$ for three different samples.
Crucible material and cooling rate have only minor effects on the electrical transport. The smaller value of the electrical resistivity at room temperature obtained for the sample grown in the Ta crucible is likely caused by the uncertainty in the geometry factor. Adjusting the geometry factor of this sample to fit the room temperature slope of $\rho(T)$ leads to an almost identical behavior when compared to the samples grown in Al$_2$O$_3$ crucibles (dashed, red line in Fig.\,\ref{res}a).  

All samples show a minimum in $\rho(T)$ at $T_{\rm min} = 15(2)$\,K (Fig.\,\ref{res}b) which can be possibly associated with the single-ion Kondo-effect. 
Comparing these data with a systematic study of $T_{\rm min}$ as a function of the impurity concentration in elemental copper\,\cite{Kjekshus1962} shows that as little as $\sim 0.01$\,\% Mn or Fe could possibly account for the observed behavior.

The residual resistivity ratios RRR = $\rho_{\rm 300\,K}$/$\rho_0$ are 8 to 10 (shown in Fig.\,\ref{res}a) and are consistent with good-quality single crystals.
(neglecting the magnetic contribution of the Kondo-effect would obviously lead to larger RRR values).
A small maximum was observed at $T = 40$\,K which is more pronounced for larger excitation currents and easily suppressed in an applied field of $\mu_0H = 0.01$\,T. 
The peak is less pronounced for the sample grown in a Ta-crucible (red, circles in Fig.\,\ref{res}), which had the largest total resistance (due to the sample geometry).
This peak is a known, frequency-dependent, artifact which is caused by a distorted low-level signal readback in the PPMS which appears in measurements of small voltage drops\,\cite{qd2000}.
It is important to point out that this does not effect the measurement at lower temperatures.

The magnetoresistance $\Delta \rho = \rho_H-\rho_0$ at $T = 1.8$\,K is positive (Fig.\,\ref{res}c, $H \parallel [1\,0\,0\,]$, $j \parallel [0\,1\,0]$). 
$\Delta\rho$ shows a sublinear field-dependence with a change of slope at $\mu_0H \approx 3$\,T.
No Shubnikov-de Haas oscillations are observed.
All samples show similar values of $\Delta \rho/\rho = 120$\,\% at $\mu_0H = 9$\,T.
The positive magnetoresistance is contrary to the expectation for strong scattering of electrons on magnetic impurities as inferred from the possible Kondo-effect.
An applied magnetic field is detrimental to the antiferromagnetic coupling of local moment and conduction electron spin. Accordingly, a negative magnetoresistance is expected, provided the magnetic contribution to the electrical resistivity is dominant. 
However, the sign of the total magnetoresistance depends on the impurity concentration: the negative, exchange contribution decreases linear with decreasing impurity concentration whereas the normal, positive part increases with decreasing impurity concentration\,\cite{Rohrer1968}.
Therefore, a positive total magnetoresistance is not unusual for low impurity concentrations (compare, e.g., with 0.0075\,\% Mn in Cu where the positive magnetoresistance is 18\,\% of the resistivity in $\mu_{0}H = 2$\,T at $T = 16$\,K and roughly temperature-independent under cooling where the negative contribution amounts to less the 10\,\% of the resistivity at $T = 1.4$\,K even after subtracting the positive contribution\,\cite{Monod1967}). 
Furthermore the large, positive metallic magnetoresistance is in accordance with a good sample quality as inferred from the emergence of de Haas-van Alphen oscillations (see below).
However, $\Delta \rho$ depends rather linear on the applied field in contrast to the quadratic behavior of a simple metal. 

\section{Temperature dependent magnetization}
\begin{figure}
\center
\includegraphics[width=0.44\textwidth]{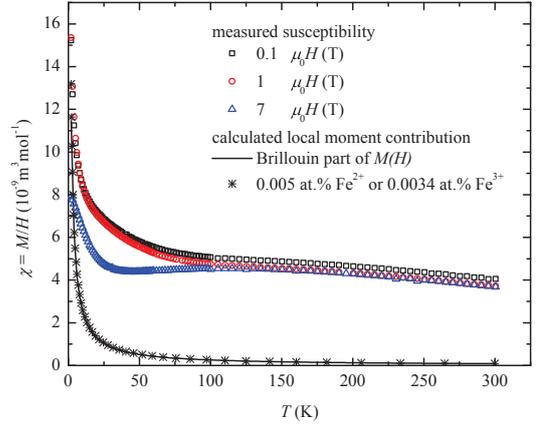}
\caption{Magnetization of \cuba~as a function of temperature ($H \parallel [1\,0\,0]$). $\chi = M/H$ increases towards low temperatures in an identical fashion for both $\mu_0H = 0.1$ and 1\,T.
The increase towards low temperatures (Curie-tail) can be well described by a contribution of diluted magnetic impurities as inferred from the $M(H)$-data (black line, calculated from the non-oscillating Brillouin part of $M(H)$ for $\mu_0H = 1$\,T). 
Assuming, e.g., Fe as possible impurity, as low as 0.005 at.\% Fe$^{2+}$ or 0.0034 at.\% Fe$^{3+}$ can account for the observed behavior (black stars, spin-only contribution).
\label{chi}}
\end{figure}

All measurements were performed with $H \parallel [1\,0\,0]$ on a sample grown in the alumina crucible (slower growth).
The temperature-dependent magnetization, $M(T)/H$, is plotted in Fig.\,\ref{chi} for applied fields of $\mu_0H = 0.1$\,T, 1.0\,T, and 7.0\,T.
Two main contributions to $M(T)/H$ can be inferred from the overall temperature dependence: 
a temperature-dependent Brillouin contribution and a temperature-independent paramagnetic term. 
The solid, black line has been calculated using the parameter obtained for the Brillouin-contribution to $M(H)$ and the temperature-dependence of the Brillouin-function (see below).
A temperature-independent contribution reflects the Pauli susceptibility of a simple metal with a value of $\chi_{\rm Pauli} \approx 3.5 \cdot 10^{-9}$\,m$^3$mol$^{-1}$.
The small field-dependence between $T = 100$ to 300\,K indicates the absence of significant amounts of ferromagnetic foreign phases (with Curie-temperatures above room-temperature).

\section{Specific heat}

\begin{figure}
\center
\includegraphics[width=0.44\textwidth]{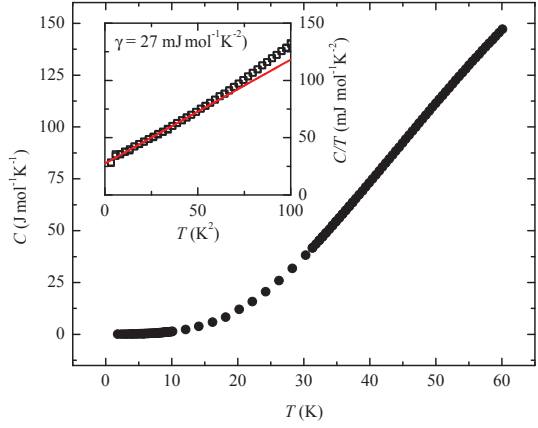}
\caption{Temperature dependent specific heat of \cuba. 
The inset shows the linear behavior of $C/T$ as a function of $T^2$, as expected for an ordinary metal, and a Sommerfeld coefficient of $\gamma = 27$\,mJ\,mol$^{-1}$K$^{-2}$ is extracted.
}
\label{hc}
\end{figure}
A sample grown in the alumina crucible (slower growth) was used for the measurement of temperature-dependent specific heat.
Figure\,\ref{hc} shows the specific heat of \cuba~for $T = 2$\,K to 60\,K. $C/T$ is plotted as a function of $T^2$ in the inset.
Electron and phonon contribution are described by $C = \gamma T + \beta T^3$ and a fit to the data is shown by a red line. 
The Sommerfeld-coefficient amounts to $\gamma = 27.0$\,mJ/(mol\,K$^{2}$).
Estimating the specific heat per atom yields $\gamma = 1.9$\,mJ/(mol-atomic\,K$^{2}$), significantly larger than the value of pure, elemental Cu ($\gamma = 0.7$\,mJ/(mol-atomic\,K$^{2})$\,\cite{Franck1961}).
A Debye-temperature of $\Theta_{\rm D} = 307$\,K was obtained from the low-temperature specific heat, close to the value observed for elemental copper ($\Theta_{\rm D} = 344$\,K)\,\cite{Franck1961}.

\section{De Haas-van Alphen oscillations}
\begin{figure}[ht!]
\center
\includegraphics[width=0.47\textwidth]{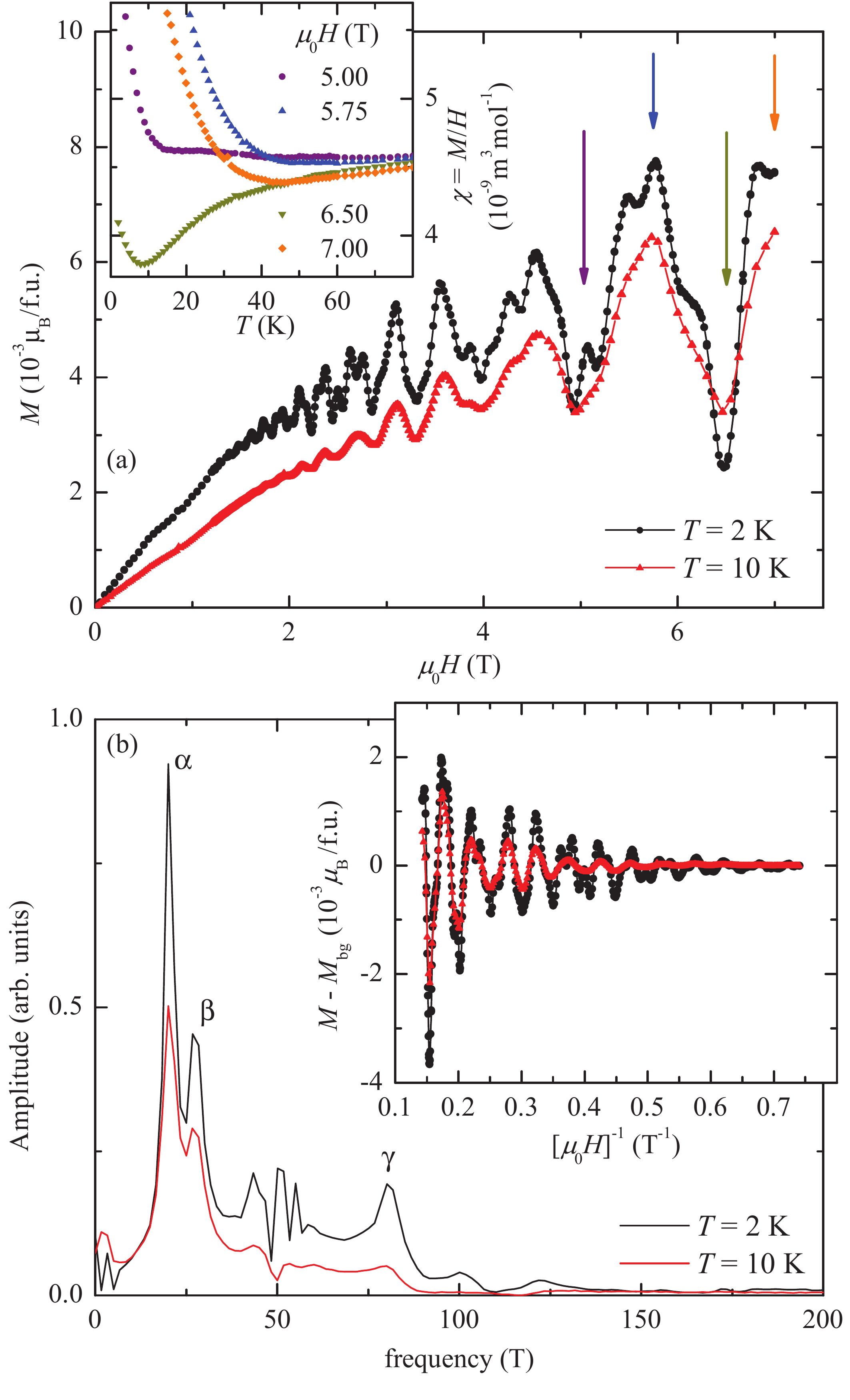}
\caption{Magnetization of \cuba~as a function of an applied field for $H \parallel [100]$.
(a) De Haas-van Alphen oscillations are observable for $\mu_0H > 1$\,T at $T = 2$\,K and for $\mu_0H > 2$\,T at $T = 10$\,K. 
$\chi(T)$ measured at different fields are shown in the inset and indicate an effect of de Haas-van Alphen oscillations up to fairly high temperatures.
The values of the applied fields correspond to minima and maxima in the $M(H)$-curves (marked by arrows).
(b) The Fourier-transformation of $M$ which was subtracted by Brillouin- and Pauli-contribution reveals several characteristic frequencies. The inset shows the oscillatory part of $M$ as function of the inverse field.
}
\label{M-H}
\end{figure}

The isothermal magnetizations at $T = 2$\,K and 10\,K show clear de Haas-van Alphen oscillations for $\mu_0H > 1$\,T and $\mu_0H > 2$\,T, respectively (Fig.\,\ref{M-H}a). 
A field-dependend background, $M_{\rm bg}$, was subtracted from $M(H)$ in order to extract the oscillatory part.
$M_{\rm bg}$ can be roughly described by a sum of a Brillouin- and a Pauli-contribution with $n = 1.0(3) \times 10^{-3}\,\mu_B$ per f.u., $g = 1.5(3)$, and $J = 2.25(50)$ (Brillouin) and $\chi = 3.5(5) \time 10^{-9}$\,m$^3$mol$^{-1}$ (Pauli).
Using these parameters and the temperature dependence of the Brillouin-function allows for a good description of the temperature dependent magnetic susceptibility over the whole investigated range from $T = 2$\,K to 300\,K (Fig.\,\ref{chi}).
The obtained oscillating $\Delta M$ is plotted in the inset of Fig.\,\ref{M-H}b as a function of the inverse magnetic field.
A Fourier-transformation reveals three sharp peaks with frequencies of 20\,T ($\alpha$), 27.5\,T ($\beta$) and 80\,T ($\gamma$). 
The amplitude of the $\alpha$-peak at $T = 10$\,K remains at $>50$\,\% of the low temperature value ($T = 2$\,K) which is consistent with a good sample quality.

De Haas-van Alphen oscillations in copper-based Kondo alloys have been observed earlier, e.g., Mn or Fe diluted in Cu\,\cite{Coleridge1971, Coleridge1972}.
The main effect of magnetic impurities is a different influence on conduction electrons of up and down spin\,\cite{Shiba1975} and in principal the de Haas-van Alphen measurements can be used to estimate the exchange energy between conduction electron spin and impurity spin\,\cite{Coleridge1970}.
However, a further analysis of the de Haas-van Alphen oscillations in \cuba~is beyond the scope of this publication.

The field-dependence of the low-temperature behavior is shown in the inset of Fig.\,\ref{M-H}a:
$\chi(T)$ strongly depends on the applied field in a very non-linear fashion which reflects the oscillatory behavior of the $M(H)$ curves.
The values of the applied fields of $\mu_0H = 5.0$\,T, 6.5\,T and $\mu_0H = 5.75$\,T, 7.0\,T correspond to minima and maxima in $M(H)$, respectively.
Therefore, the temperature-dependent magnetization of \cuba~is affected by de Haas-van Alphen oscillations up to temperatures as high as 50\,K.

\section{Summary}
\cuba~offers straight forward single crystal growth, rather simple cubic crystal structure with a large unit cell, and quality sufficient to observe de Haas-van Alphen oscillations. 
The minimum in electrical resistivity is reminiscent of the behavior observed for dilute magnetic impurities in elemental copper, e.g., iron in copper\,\cite{Kjekshus1962,Franck1961}. 
In accordance, temperature- and field dependence of the magnetization indicate the presence of magnetic impurities with a concentration of the order of $\sim$\,0.01\,at.\%.
Whereas \cuba~appears to be a relatively well behaved, simple metal, a systematic introduction of iron or other magnetic impurities in \cuba~could allow for the study of the impurity Kondo effect in binary compounds and, perhaps, an improved understanding of local moment formation in metals by elucidating minor or major differences to pure copper.

% References
%
% Following citation commands can be used in the body text:
% Usage of \cite is as follows:
%   \cite{key}          ==>>  [#]
%   \cite[chap. 2]{key} ==>>  [#, chap. 2]
%   \citet{key}         ==>>  Author [#]

\section{Acknowledgment}
This work was supported by the U.S. Department of Energy, Office of Basic Energy Science, Division of Materials Sciences and Engineering. The research was performed at the Ames Laboratory. Ames Laboratory is operated for the U.S. Department of Energy by Iowa State University under Contract No. DE-AC02-07CH11358.

% \bibliographystyle{model1-num-names1}
% \bibliography{zitate}

\end{document}